\documentclass[a4paper,11pt]{article}
\pdfoutput=1
\usepackage{jinstpub} 
\usepackage{subcaption}
\usepackage{xcolor}
\usepackage{hyperref}
\usepackage{lineno}
\usepackage{soul}
\usepackage{orcidlink}
\usepackage{upgreek}

\title{\boldmath Muon tracking in a LiquidO opaque scintillator detector}

\collaboration{
LiquidO Collaboration}

\newcommand{\Londrina}{Departamento de F\'isica, 
	Universidade Estadual de Londrina, 
	Londrina,
	Brazil}

\newcommand{\PUCR}{Department of Physics, 
	Pontif\'icia Universidade Cat\'olica do Rio de Janeiro, 
	Rio de Janeiro, Brazil}

\newcommand{\Queens}{Department of Physics, 
	Engineering Physics \& Astronomy, 
	Queen's University, Kingston, 
	Canada}

\newcommand{\Prague}{Institute of Particle and Nuclear Physics,
	Charles University,
	Prague, 
	Czech Republic}

\newcommand{\IJCLabSaclay}{Universit\'e Paris-Saclay, 
	CNRS/IN2P3, 
	IJCLab, 
	Orsay, 
	France}

\newcommand{\LPtwoI}{Universit\'e de Bordeaux, 
	CNRS, 
	LP2I Bordeaux, 
	Gradignan, 
	France}

\newcommand{\SUBA}{Nantes Universit\'e, 
	IMT-Atlantique, 
	CNRS, 
	Subatech,
	Nantes, 
	France}
	
\newcommand{\CPPM}{Universit\'e de Aix Marseille, 
	CNRS, 
	CPPM, 
	Marseille, 
	France}

\newcommand{\LNCA}{LNCA Underground Laboratory, 
	CNRS, 
	EDF Chooz Nuclear Reactor, 
	Chooz, France}

\newcommand{\FerraraUni}{Dipartimento di Fisica e Scienze della Terra, 
	Universit\`{a} di Ferrara, 
	Ferrara, 
	Italy}
\newcommand{\FerraraINFN}{INFN, 
	Sezione di Ferrara, 
	Ferrara, 
	Italy}

\newcommand{\Padovauni}{Dipartimento di Fisica e Astronomia, 
	Universit\`{a} di Padova, 
	Padova, Italy}
\newcommand{\Padova}{INFN, 
	Sezione di Padova, 
	Padova, Italy}

\newcommand{\MainzA}{Johannes Gutenberg-Universit\"{a}t Mainz,
	Institut f\"{u}r Physik, 
	Mainz, Germany} 
\newcommand{\MainzB}{Johannes Gutenberg-Universit\"{a}t Mainz,
	Detektorlabor, Exzellenzcluster PRISMA$^+$,
	Mainz, Germany} 	

\newcommand{\TechnicoIBB}{iBB, 
	Instituto Superior Tecnico, 
	Universidade de Lisboa,
	Lisbon, 
	Portugal}
    
\newcommand{\TechnicoIDMEC}{IDMEC, 
	Instituto Superior Tecnico, 
	Universidade de Lisboa,
	Lisbon, 
	Portugal}

\newcommand{\TechnicoCtwoTN}{CT2N, 
	Instituto Superior Tecnico, 
	Universidade de Lisboa,
	Lisbon, 
	Portugal}

\newcommand{\CIEMAT}{CIEMAT, 
	Centro de Investigaciones Energ\'{e}ticas, Medioambientales y Tecnol\'{o}gicas, 
	Madrid, Spain}

\newcommand{\UniZar}{Centro de Astropart\'{\i}culas y F\'{\i}sica de Altas Energ\'{\i}as (CAPA),
 	Universidad de Zaragoza, 
	Zaragoza, Spain}

\newcommand{\DIPC}{Donostia International Physics Center, 
	Basque Excellence Research Centre, 
	San Sebasti\'an/Donostia, 
	Spain}

\newcommand{\RCNS}{RCNS, 
	Tohoku University, 
	Sendai, Japan}

\newcommand{\Sussex}{Department of Physics and Astronomy, 
	University of Sussex, 
	Brighton, 
	United Kingdom}

\newcommand{\ImpCol}{Department of Chemistry, 
	Imperial College London, 
	London, 
	United Kingdom}
	
\newcommand{\RAL}{Rutherford Appleton Laboratory, 
	Didcot,
	Oxford,
	United Kingdom}

\newcommand{\UCI}{Department of Physics and Astronomy, 
	University of California at Irvine, 
	Irvine, 
	CA, 
	USA}

\newcommand{\UniPennPhys}{Department of Astronomy and Astrophysics, 
	Pennsylvania State University, 
	University Park, 
	PA,
	USA}
\newcommand{\UniPennAstro}{Department of Physics, 
	Pennsylvania State University, 
	University Park, 
	PA,
	USA}

\newcommand{\UniMichigan}{Department of Nuclear Engineering and Radiological Sciences,
 	University of Michigan, 
	Ann Arbor, 
	MI,
	USA}
					
\newcommand{\BNL}{Brookhaven National Laboratory, 
	Upton, 
	NY,
	USA}

\author[z]{J.\,Apilluelo,}
\author[b]{L.\,Asquith,}
\author[b]{E.\,F.\,Bannister,}
\author[k\alpha]{N.\,P.\,Barradas,}
\author[b]{C.\,L.\,Baylis,}
\author[p]{J.\,L.\,Beney,}
\author[k\beta]{M.\,Berberan\,e\,Santos,}
\author[p]{X.\,de\,la\,Bernardie,}
\author[b]{T.\,J.\,C.\,Bezerra \orcidlink{0000-0002-0424-7903},}
\author[p]{M.\,Bongrand,} 
\author[q]{C.\,Bourgeois,}
\author[q]{D.\,Breton,}
\author[n]{J.\,Busto,}
\author[q,c]{A.\,Cabrera \orcidlink{0000-0001-5713-3347},}
\author[p]{A.\,Cadiou,}
\author[l]{E.\,Calvo,}
\author[b]{M.\,de\,Carlos\,Generowicz,}
\author[f]{E.\,Chauveau,}
\author[b]{B.\,J.\,Cattermole,}
\author[h]{M.\,Chen,}
\author[i]{P.\,Chimenti,} 
\author[x\alpha,x\beta]{D.\,F.\,Cowen,}
\author[b]{S.\,Kr.\,Das,}
\author[r\alpha]{S.\,Dusini \orcidlink{0000-0002-1128-0664},} 
\author[b]{A.\,Earle,}
\author[k\alpha]{M.\,Felizardo,}
\author[i]{C.\,Frigerio\,Martins,}
\author[z]{J.\,Gal\'an,}
\author[z]{J.\,A.\,Garc\'ia,}
\author[b]{A.\,Gibson-Foster,}
\author[m\alpha]{C.\,Girard-Carillo,}
\author[b]{W.\,C.\,Griffith,}
\author[u]{J.\,J.\,G\'omez-Cadenas,}
\author[p]{M.\,Guiti\`ere,}
\author[p]{F.\,Haddad,}
\author[b]{J.\,Hartnell \orcidlink{0000-0002-1744-7955},} 
\author[d]{A.\,Holin,}
\author[z]{I.\,G.\,Irastorza,}
\author[a]{I.\,Jovanovic \orcidlink{0000-0003-0573-3150},}
\author[k\alpha]{A.\,Kling,}
\author[m\alpha]{L.\,Koch \orcidlink{0000-0002-2966-7461},}
\author[b]{P.\,Lasorak,}
\author[q,c]{J.\,F.\,Le\,Du,}
\author[p]{F.\,Lefevre,}
\author[q]{P.\,Loaiza,}
\author[b]{J.\,A.\,Lock,}
\author[z]{G.\,Luz\'on,}
\author[q]{J.\,Maalmi,}
\author[j]{J.\,P.\,Malhado,}
\author[e\alpha,e\beta]{F.\,Mantovani,}
\author[k\alpha]{J.\,G.\,Marques,}
\author[f]{C.\,Marquet,} 
\author[z]{M.\,Mart\'inez,}
\author[x\beta]{J.\,T.\,Moffat,}
\author[l]{D.\,Navas-Nicol\'as \orcidlink{0000-0002-2245-4404},}
\author[t]{H.\,Nunokawa,} 
\author[g]{J.\,P.\,Ochoa-Ricoux \orcidlink{0000-0001-7376-5555},} 
\author[k\beta]{T.\,Palmeira,}
\author[l]{C.\,Palomares,} 
\author[d]{D.\,Petyt,}
\author[p]{P.\,Pillot,}
\author[f]{A.\,Pin,}
\author[b]{J.\,C.\,C.\,Porter,} 
\author[f]{M.\,S.\,Pravikoff \orcidlink{0000-0002-7088-4126},}
\author[d]{S.\,Richards,}
\author[k\beta]{N.\,Rodrigues,}
\author[f]{M.\,Roche,}
\author[y]{R.\,Rosero,}
\author[s]{B.\,Roskovec,}
\author[z]{M.\,L.\,Sarsa,}
\author[m\beta]{S.\,Schoppmann \orcidlink{0000-0002-7208-0578},}
\author[r\alpha,r\beta]{A.\,Serafini,}
\author[d]{C.\,Shepherd-Themistocleous,}
\author[b]{W.\,Shorrock \orcidlink{0000-0002-7221-1910},}
\author[k\gamma]{M.\,Silva,}
\author[q]{L.\,Simard,}
\author[u]{S.\,R.\,Soleti,}
\author[p]{D.\,Stocco,}
\author[e\alpha,e\beta]{V.\,Strati,}
\author[p]{J.\,S.\,Stutzmann,}
\author[v]{F.\,Suekane,}
\author[b]{N.\,Tuccori \orcidlink{0000-0002-2868-5887},}
\author[l]{A.\,Verdugo,}
\author[p]{B.\,Viaud,}
\author[m\alpha]{S.\,M.\,Wakely \orcidlink{0000-0002-2919-8159},}
\author[m\alpha]{A.\,Weber \orcidlink{0000-0002-8222-6681},}
\author[x\beta]{G.\,Wendel \orcidlink{0000-0002-2440-9391},}
\author[a]{A.\,S.\,Wilhelm \orcidlink{0000-0002-0664-0477},}
\author[b,d]{A.\,W.\,R.\,Wong,}
\author[y]{M.\,Yeh,}
\author[p]{F.\,Yermia}
\affiliation[a]{\UniMichigan} 
\affiliation[b]{\Sussex} 
\affiliation[c]{\LNCA} 
\affiliation[d]{\RAL} 
\affiliation[e\alpha]{\FerraraINFN} 
\affiliation[e\beta]{\FerraraUni} 
\affiliation[f]{\LPtwoI} 
\affiliation[g]{\UCI} 
\affiliation[h]{\Queens} 
\affiliation[i]{\Londrina} 
\affiliation[j]{\ImpCol} 
\affiliation[k\alpha]{\TechnicoCtwoTN} 
\affiliation[k\beta]{\TechnicoIBB} 
\affiliation[k\gamma]{\TechnicoIDMEC} 
\affiliation[l]{\CIEMAT} 
\affiliation[m\alpha]{\MainzA} 
\affiliation[m\beta]{\MainzB} 
\affiliation[n]{\CPPM} 
\affiliation[p]{\SUBA} 
\affiliation[q]{\IJCLabSaclay} 
\affiliation[r\alpha]{\Padova} 
\affiliation[r\beta]{\Padovauni} 
\affiliation[s]{\Prague} 
\affiliation[t]{\PUCR} 
\affiliation[u]{\DIPC} 
\affiliation[v]{\RCNS} 
\affiliation[x\alpha]{\UniPennPhys} 
\affiliation[x\beta]{\UniPennAstro} 
\affiliation[y]{\BNL} 
\affiliation[z]{\UniZar} 

\emailAdd{LiquidO-Contact-L@in2p3.fr}

\abstract{LiquidO is an innovative radiation detector concept. The core idea is to exploit stochastic light confinement in a highly scattering medium to self-segment the detector volume. %
In this paper, we demonstrate event-by-event muon tracking in a LiquidO opaque scintillator detector prototype. The detector consists of a 30~mm cubic scintillator volume instrumented with 64 wavelength-shifting fibres arranged in an 8$\times$8 grid with a 3.2~mm pitch and read out by silicon photomultipliers. A wax-based opaque scintillator with a scattering length of approximately 0.5~mm is used. The tracking performance of this LiquidO detector is characterised with cosmic-ray muons and the position resolution is demonstrated to be 450~$\upmu$m per row of fibres. %
These results highlight the potential of LiquidO opaque scintillator detectors to achieve fine spatial resolution, enabling precise particle tracking and imaging. %
}

\keywords{Particle tracking detectors, scintillators and scintillating fibres and light guides, scintillation and light emission processes (solid, gas and liquid scintillators)}

\begin{document}
\maketitle
\flushbottom

\section{Introduction}
\label{sec:intro}
LiquidO represents a novel approach to radiation detection~\cite{liquido0}. The core of this innovative concept is to constrain light near its emission point through stochastic confinement within highly scattering opaque media~\cite{minipaper}, and then rapidly collect it in situ. %
For this process to be effective, materials that have a short scattering length and low absorption of the light are required. The scattering causes photons---whether produced by scintillation or other light-emission processes---to undergo a random walk around their point of production, resulting in the ``LiquidO effect'', which confines light to a spherical region around each emission point. %
A lattice of wavelength-shifting (WLS) fibres collects and directs the light to photosensors, typically silicon photomultipliers (SiPMs). Consequently, the topology of the energy deposition is preserved, and the detection volume is effectively self-segmented. This unique characteristic grants LiquidO-based radiation detectors high-resolution imaging capability, enabling precise reconstruction of event topologies and accurate particle identification on an event-by-event basis.

In contrast, traditional detectors using transparent light-emitting media that seek topological information either surround the detection volume with photosensors~\cite{sno, superk, borexino, doublechooz, juno}, or physically segment the sensitive volume~\cite{minos, nova, t2k, SOLID, superfgd}. %
The former method does not usually yield high-resolution topological information, while the latter achieves this at the expense of adding more inactive material to the detector volume and increasing light-collection inefficiencies, radioactive contaminants and manufacturing complexity. %
Some designs integrate a fibre lattice within a transparent scintillator volume~\cite{scibath, finesse}, but they do not benefit from the stochastic light confinement of a highly scattering opaque medium, which keeps light localised near its point of production. %

The development of the novel LiquidO technology is led by the international LiquidO collaboration~\cite{liquido_project}. Notable projects within the LiquidO framework include the large antineutrino detectors AntiMatter-OTech/CLOUD~\cite{amotech_EIC, cabrera_nutel} and SuperChooz~\cite{cabrera2019}. Additionally, applications that could benefit from the use of a LiquidO detector encompass positron emission tomography (PET)~\cite{lpet_anatael}, geoneutrino detection~\cite{geoneutrino}, high-energy electromagnetic calorimetry~\cite{cal_hull}, neutrinoless double beta decay searches~\cite{nudoubt}, and astrophysical explorations~\cite{cocoa}. %

Previous LiquidO prototypes have established the foundational principles of this detection technique. Mini-LiquidO demonstrated stochastic light confinement around point-like MeV-scale energy depositions while operating in a regime of low detected light (10~photoelectrons/MeV) but high sensitivity~\cite{minipaper}. Additional work with a 1-litre, 32-fibre detector investigated LiquidO’s vertex reconstruction capabilities for point-like light emissions, demonstrating sub-centimetre precision~\cite{LIME}. %

This article describes the characterisation of a 64-fibre LiquidO opaque scintillator detector prototype, referred to as the Cube detector, through the detection and reconstruction of cosmic-ray muon tracks. This is made possible by an unprecedented number of readout channels (128) for any LiquidO detector to date. We achieve a sub-millimetre, one-dimensional position resolution, demonstrating the potential of LiquidO detectors for applications demanding precise tracking. In particular, for muon imaging applications~\cite{PROCUREUR2018169}, LiquidO offers a promising alternative to conventional systems based on segmented plastic scintillator bars~\cite{anghel2015cript, luo2022lumis} or densely packed scintillating fibres~\cite{gscan1, Mahon20180048}. %

Moreover, we confirm on an event-by-event basis the presence of stochastic light confinement in a highly scattering opaque medium---a defining feature of the LiquidO detection principle--- in a regime complementary to Mini-LiquidO~\cite{minipaper}, with detected light levels roughly twenty times higher. %

The publication contents are organised as follows. %
Section~\ref{sec:setup} details the Cube prototype design, including information on the scintillators used, the setup designed for muon detection, and the readout system. Section~\ref{sec:daq} presents the data acquisition method, and Section~\ref{sec:selection} describes the selection criteria for muon events. Section~\ref{sec:tot} provides a method to correct the nonlinearity of the light measurement that arises from the implementation of the readout system. Section~\ref{sec:displays} shows experimental event displays that visually highlight muon tracks and the difference between using an opaque and a transparent scintillator. Section~\ref{sec:resolution} contains the experimental results of the position resolution of the Cube. In Section~\ref{sec:imaging} we give context for the achieved resolution, comparing it with state-of-the-art muon-imaging scintillator-based systems and discuss the potential of an optimised LiquidO detector. Finally, Section~\ref{sec:conclusion} summarises the work.

\section{The 64-fibre Cube detector}
\label{sec:setup}
\begin{figure}[htbp]
\centering
\includegraphics[width=\textwidth]{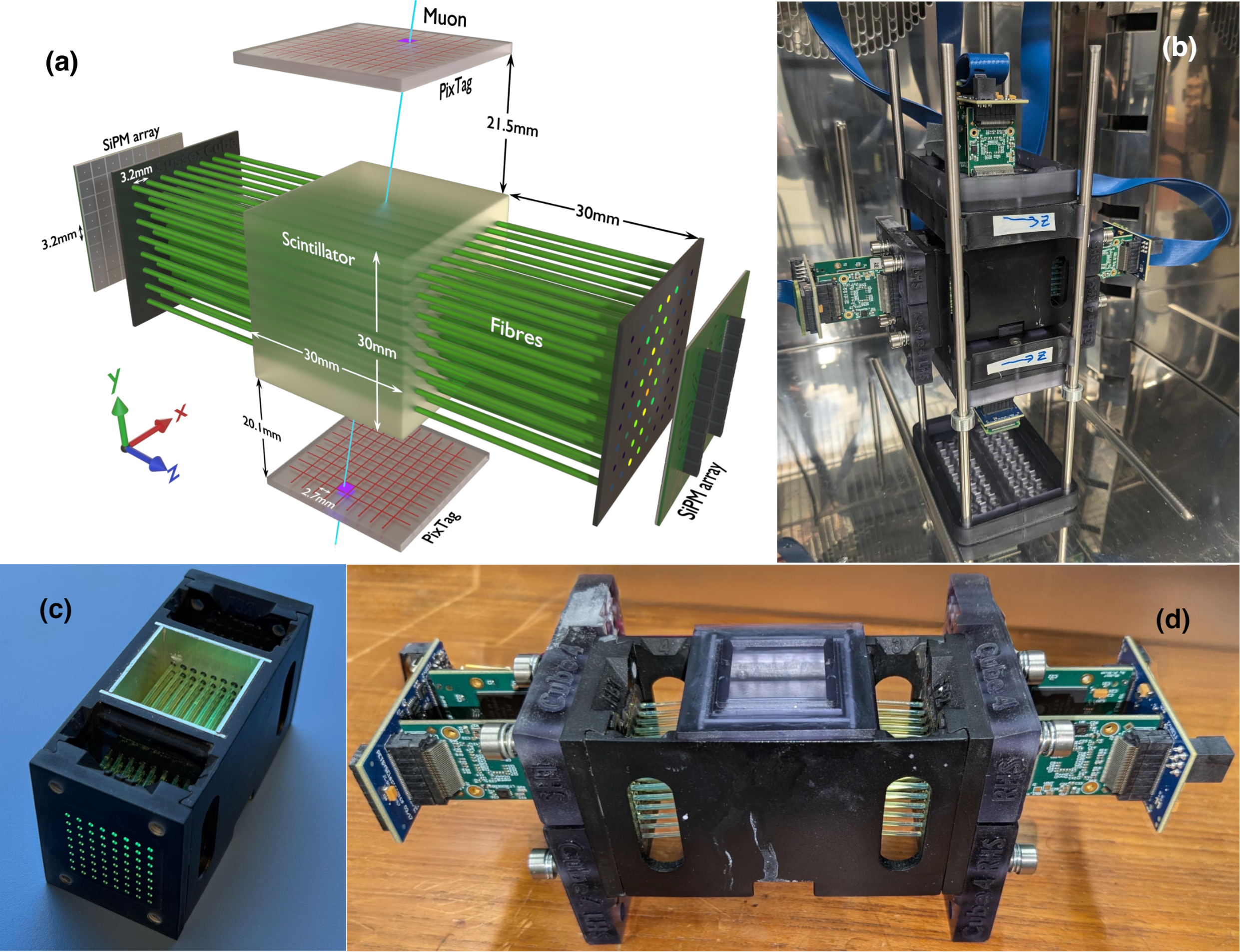}
\caption{Images of the 64-fibre Cube detector. (a) A three-dimensional graphical representation of the detector displaying the internal components. %
(b) A photo of the setup, composed of the instrumented detector in between the two pixelated muon taggers, placed above and below the detector and enclosed in a black housing. (c) A photo of the detector not filled with scintillator, showing the fibre lattice. (d) A photo of the instrumented detector alone, with the readout ASICs and SiPMs attached and the lid on. %
\label{fig:setup_diagram}}
\end{figure}

Photos and a diagram of the 64-fibre Cube prototype detector, are shown in Figure~\ref{fig:setup_diagram}. %
The skeleton (Figure~\ref{fig:setup_diagram}c) is 3D printed using Elegoo ABS-Like Resin, providing a volume of 30$\times$30$\times$30~mm$^3$ to be filled with scintillator. The interior walls are made of aluminium plates which are highly reflective for the wavelengths emitted by the scintillator~\cite{AlRefl}. The top part of the Cube is enclosed using a lid (Figure~\ref{fig:setup_diagram}d), which has an aluminium plate on the side facing the scintillator. %
Saint Gobain's BCF-91A WLS double-clad fibres (1~mm diameter)~\cite{SaintGobain2019} are threaded through the skeleton with a 3.2~mm spacing and sealed with a liquid insulating rubber to prevent leaks. The fibres are secured in place in the end blocks with a two-part epoxy pigmented black. The fibre ends are finished using a fly cutter, which provides a smooth end and improves photon transmission between the fibre and SiPM, which are air-coupled. The BCF-91A fibres comprise a polystyrene core surrounded by an acrylic layer and an outer fluor-acrylic layer, providing a minimum trapping efficiency of 5.6\% per fibre end, as quoted in the datasheet. The WLS dye absorbs light in the 350–470~nm range, with an absorption peak at 420~nm, and emits over 460–600~nm, with an emission peak at 490~nm. %
The choice of 1 mm fibres was primarily guided by practical considerations: they are mechanically robust, allowing repeated filling and handling of the Cube, and they absorb scintillation photons efficiently. %

\begin{figure}[htbp]
\centering
\includegraphics[width=\textwidth]{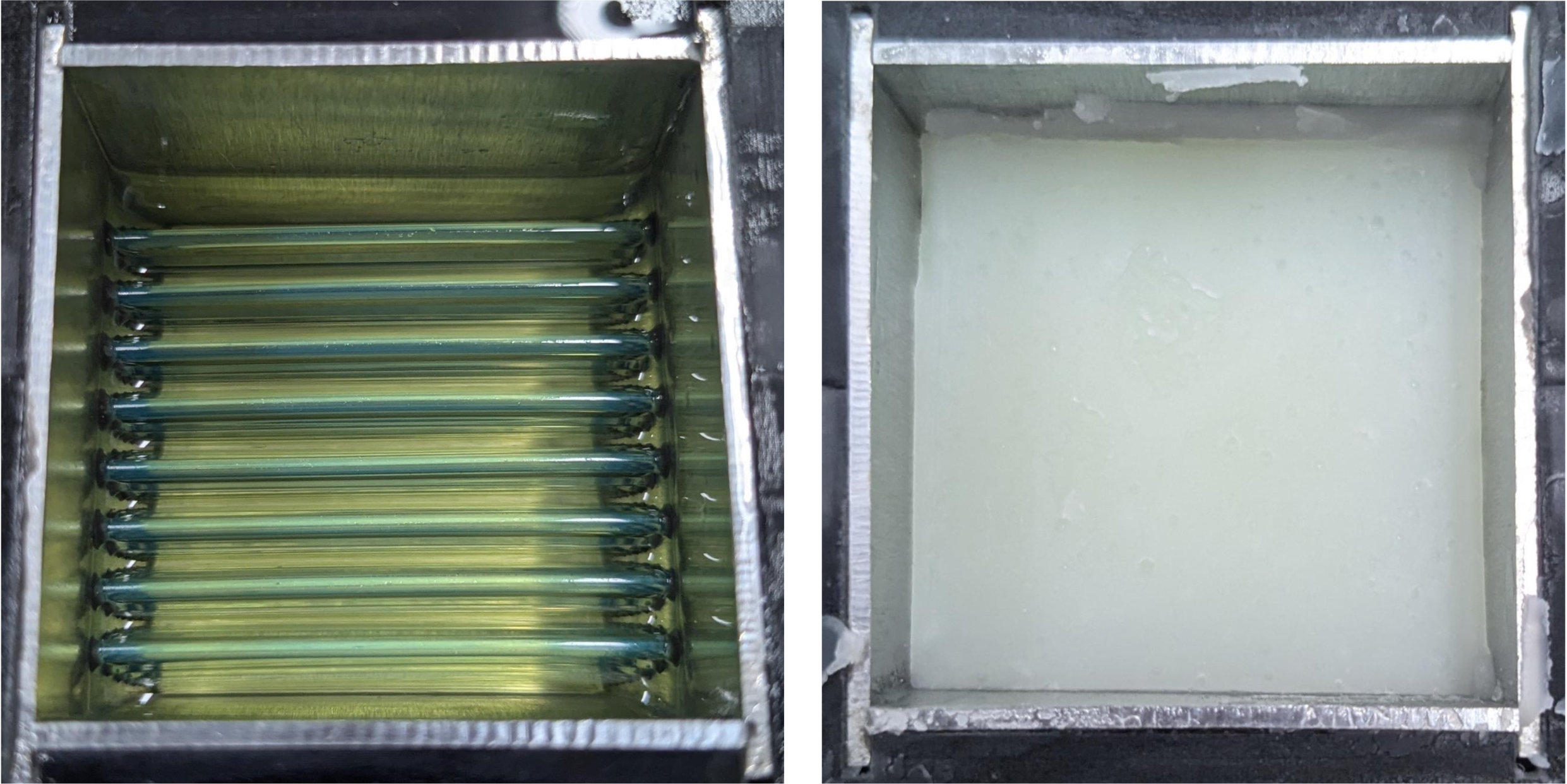}
\caption{Photos of the interior of the 64-fibre Cube detector filled with the scintillators used: the liquid transparent mixture (left) and the opaque NoWaSH (right). %
\label{fig:scints_photos}}
\end{figure}

We tested both transparent and opaque scintillators in the Cube detector. The opaque scintillator used is a derivative of the wax-based opaque scintillators named ``NoWaSH''~\cite{buck2019_nowash}. %
The version of NoWaSH used is designed with approximately 0.5~mm scattering length~\cite{stefan_thesis} so that the opacity of the medium confines the majority of the light to a region on the scale of the fibre pitch. In this formulation, the NoWaSH is produced by mixing 15~wt.\% wax with a base of transparent liquid scintillator, which comprises LAB, 1~wt.\% PPO, and 0.03~wt.\% POPOP\@. %
This liquid scintillator has an absorption length on the order of meters at its peak emission wavelength and an intrinsic light yield of about 9000~photons/MeV~\cite{buck2019_nowash}. The two fluors in the scintillator base allow for repeated emission/absorption to shift the scintillator photons to better match the absorption spectrum of the fibres. %
The LAB+PPO+POPOP transparent scintillator base is also used to acquire data, enabling a comparison between opaque and transparent scintillators. The only difference between the two scintillators is the presence of wax in the NoWaSH\@. 
Figure~\ref{fig:scints_photos} includes photos of the detector filled with transparent and opaque scintillators. %

The detection of cosmic muons is assisted by two pixelated cosmic muon taggers (PixTags), placed above and below the Cube (see Figure~\ref{fig:setup_diagram}a,~\ref{fig:setup_diagram}b). These supplementary detectors are used to tag charged particles crossing the Cube. Each PixTag contains 64 scintillator pixels, which are directly read out by an array of SiPMs. The scintillator arrays are produced from transparent polystyrene slabs doped with 1~wt.\% PPO and 0.03~wt.\% POPOP~\cite{minos}. Grooves are cut into the 3~mm thick scintillator and filled with red resin to create optically separated squares of 3.2$\times$3.2~mm$^2$, matching the dimensions of the SiPMs. The scintillator is flush with the SiPM array, such that each pixel is paired with a SiPM from the 8$\times$8 array. A plastic housing encloses the PixTags to prevent light leakage and reduce external noise. The PixTags are positioned about 20~mm above and below the opaque scintillator volume (see Figure~\ref{fig:setup_diagram}a).

The SiPM arrays used to read out the Cube and the two PixTags are Hamamatsu S13361-3050AE-08 with 64 channels in an 8$\times$8 grid~\cite{hamamatsu_mppc}. Each SiPM has an effective photosensitive area of 3$\times$3~mm$^2$ and a photon detection efficiency of 38\% at the fibre peak emission wavelength of 490~nm, as specified in the datasheet for an overvoltage of 3~V. %
The electronics front-end modules (FEMs) plugged into each SiPM array are held in place with spring-loaded screws to the fibre end blocks to maintain consistent pressure to the face of the SiPMs (Figure~\ref{fig:setup_diagram}d). %
The four 64-SiPM arrays used in this setup, two for the two ends of the Cube and one each for the two PixTags, are read out by the commercially available TOFPET2 ASICs developed by PETsys Electronics~\cite{bugalho2019_petsys, rolo2013_petsys}. These low-power, 64-channel chips read out and digitise signals from the SiPMs, streaming data that includes the channel number, timestamp, and charge estimate of each event. The four PETsys FEMs are connected to a PETsys FEB/D-1k module, which performs system-wide online coincidence between the FEMs and sends the data from the ASICs to the data acquisition (DAQ) computer. The FEMs are equipped with temperature sensors, allowing the regular monitoring of the temperature of the ASICs and the SiPM arrays. The temperature of the two SiPM arrays reading out the sides of the Cube were 23.5$^\circ$C and 25.0$^\circ$C, with a stability of 0.5$^\circ$C over the entire data-taking period. The entire setup was operated inside a light-tight climate chamber (see Figure~\ref{fig:setup_diagram}b) with a stable air temperature of 18.0$\pm$0.3~$^\circ$C.

\section{Data acquisition and grouping methods}
\label{sec:daq}
The SiPM array connected to the left (right) side of the Cube has a breakdown voltage of 51.30 V (51.90 V) and is operated with an overvoltage of 3.50 V (3.75 V). Each array of 64 SiPMs is supplied with an independent bias voltage, allowing fine adjustment of the operating point. The small difference in overvoltage is intentionally set to equalize the gain between the two arrays, ensuring a more uniform response across all 128 SiPMs and enabling the use of a common trigger threshold for all channels. %
The trigger threshold is set to detect pulses generated by at least two photoelectrons (p.e.). This low threshold is selected as a trade-off between the light collection efficiency and dark noise. The SiPM arrays connected to the PixTags have a 51~V breakdown voltage and are operated with 3~V overvoltage. A trigger threshold of 5 p.e.\ is set in the PixTags to reduce the likelihood of noise generating a signal. By utilising the online coincidence trigger of the PETsys acquisition system, only pulses occurring in coincidence---where a signal above the threshold is detected for a PixTag and at least one other detector (either the other PixTag or either side of the Cube)---are recorded. Given that muons produce large and clean signals in the PixTags, it is likely that corresponding signals in the other detectors are caused by the same muon. Therefore, this online coincidence trigger pre-filters for cosmic muons and rejects the large majority of noise events occurring on the two sides of the Cube, where a low threshold is used. This setup gives a low overall trigger rate of 0.4~Hz, corresponding to only 4~MB of data per day.

For each ``hit'', defined as a single digitised pulse from a SiPM, we record the time at which the signal went above threshold, the channel number of the SiPM in which the hit occurred, and an estimate of the number of p.e.\ that generated the hit. The TOFPET2 ASIC has the capability of performing the p.e.\ measurement in terms of either charge integration (QDC) or Time-over-Threshold (ToT). The charge integration of the TOFPET2 ASIC is unable to resolve signals below about 20 p.e. The signal from each SiPM in our detector ranges from approximately 1 to 100 p.e.\ per pulse. Therefore, we have chosen to use the ToT measurement. It is important to note that ToT measurements exhibit a non-linear response. For this reason, a linearisation method, discussed in Section~\ref{sec:tot}, is applied offline. %
Events, defined as a group of hits, are formed offline by grouping hits from all 256 SiPM channels using a time window of 50~ns. %

\section{Offline muon selection}
\label{sec:selection}
The online coincidence trigger discussed in Section~\ref{sec:daq} filters out the majority of background and noise events. We also apply an offline selection to obtain a sample of clean muon events, in particular using the independent PixTags and several timing cuts. All our cuts are applied to both the transparent and opaque datasets to avoid any potential biases. %

We first select only particles passing through the Cube and both PixTags. The selection imposes the condition that there must be hits in both PixTags within a 6~ns time window and above the energy threshold of background events. %
This 6~ns window was chosen based on the observed distribution of time differences between first hits in the PixTags. It is significantly larger than the about 100~ps transit time of a muon through the detector as it accounts for the scintillator response and no offline timing calibration was performed. %
This enables the selection of through-going particle events---most likely muons---by relying on the signals from the two PixTags only. %

To ensure the time structure of events was as expected, the hits recorded by both PixTags had to be present in the first 10~ns of an event, and the first hit from the SiPMs reading out the Cube had to be recorded within 5~ns of the PixTag hits. This selection condition rejected 2\% of the previously selected events. %

The resulting event rate at this stage is 10.4 muons/hour. %
This rate is consistent with the analytical prediction based on the flux of muons at the earth's surface and the convolution of the angular acceptance of the two PixTags and their geometry~\cite{GRThomas_1972}. Over the various weeks-long acquisitions the event rate was observed to be stable with variation less than 9\% every 12 hours of acquisition, consistent with the expected statistical uncertainty. %

An additional selection is applied to the muon sample for the position resolution analysis described in Section~\ref{sec:resolution}. %
To eliminate particles that partially cross the detector and trigger due to a noise hit in the other PixTag, events where an entire row of SiPMs in one of the arrays connected to the the Cube fail to register signals are excluded. This step removes approximately 3\% of remaining candidate muons. %

To filter out non-tracklike events that have passed the previous selection cuts, such as those caused by multiple particles (e.g., delta-rays produced by a passing muon) crossing the scintillator volume simultaneously, we apply two additional cuts. The first cut removes events with multiple hits in one PixTag, except when the extra hits are adjacent to the highest-energy pixel. The second cut removes events with unusual track widths in the detector, suggesting more than one particle is depositing energy. We calculate the average spread of scintillation light in a row of fibres for each event, defined as the standard deviation with respect to the weighted average of the signals in each row. Afterwards, the overall average and standard deviation of these widths for each dataset are found. Events with widths that deviate more than 3~$\upsigma$ from the average are discarded. These two cuts remove about 6\% and 8\% of events respectively. %

The reconstruction technique described in Section~\ref{sec:resolution} introduces a bias for events near the detector's borders, causing tracks to appear further from the border than they actually are. To avoid this, we select muons that pass through the detector central region and reject the ones that travel vertically close to the borders of the scintillator in the x-direction (see Figure~\ref{fig:setup_diagram} for axis reference). This is done by examining the positions of the largest signals in the two PixTags during each event. Events that in both PixTags have their maximum-energy hit in the two pixels at the far left or right on the x-axis are removed. This final cut removes about 37\% of muons.

Overall, this additional selection removes around 47\% of the muon candidates previously identified. After this selection, the datasets consist of around 3500 muons for each scintillator used to fill the Cube, acquired over time periods of approximately four weeks each.

\section{Linearisation of the Time-over-Threshold signal}
\label{sec:tot}
Time-over-Threshold is a power-efficient and cost-effective method to estimate the number of p.e.\ that generates a signal. It is used in PET and high-energy physics applications. A drawback to using ToT is that, in typical scintillator detectors, the exponential decay of the pulse causes the time it remains above the threshold to increase logarithmically with the number of p.e.\ generating the pulse. This introduces a strong nonlinearity, implying that the higher the number of p.e.\ generating the signal, the less accurate ToT is. %
As such, when using ToT as a measurement of the size of a signal, this nonlinearity has to be addressed. %

\begin{figure}[htbp]
\centering
\includegraphics[width=\textwidth]{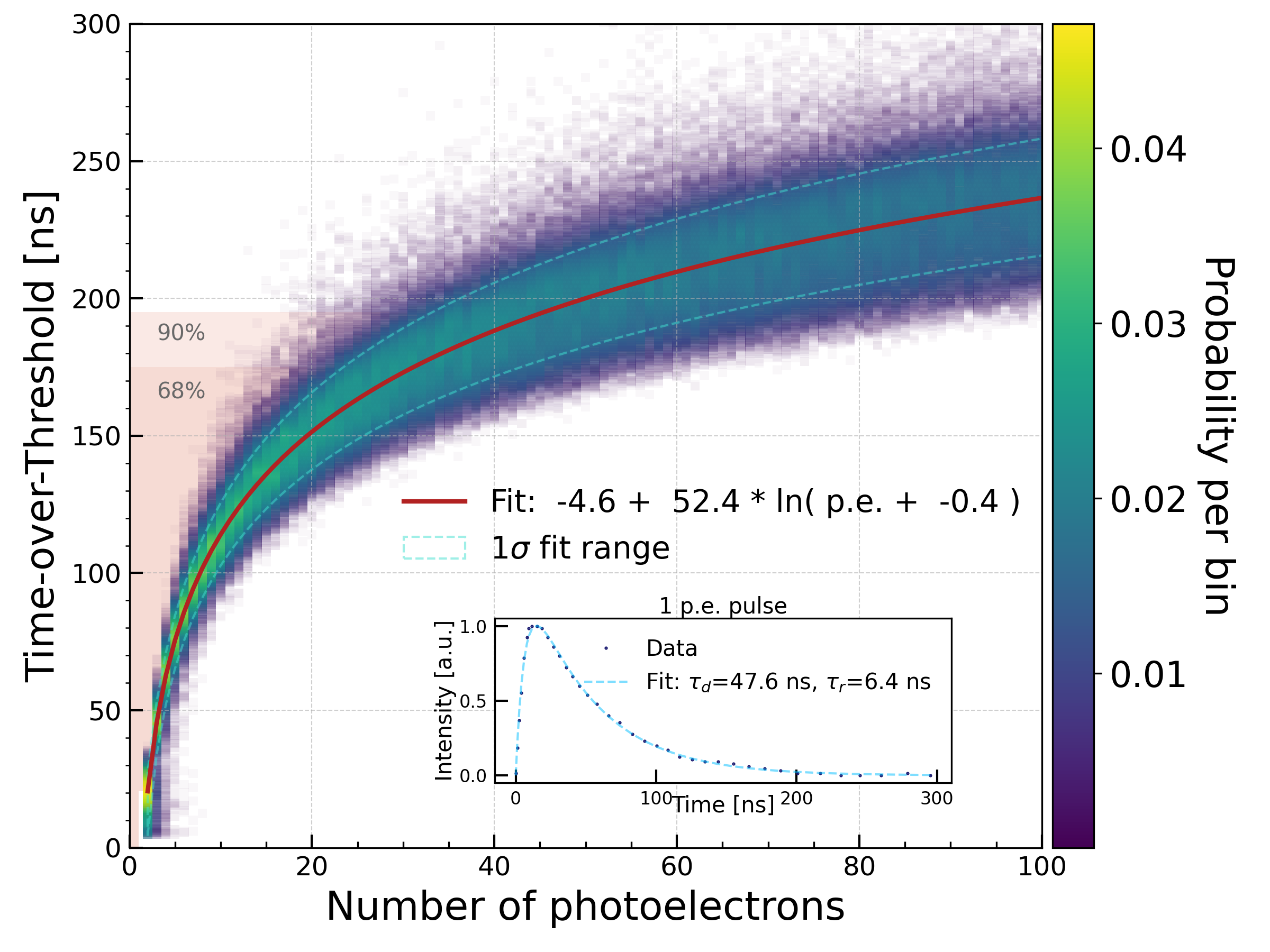}
\caption[Probability map of ToT values as a function p.e.]{Probability map of ToT values as a function of the number of p.e. For each p.e.\ value, pulses replicating experimental signals are generated using a Monte-Carlo simulation, and used to calculate the corresponding ToT values and the ToT probability distribution. The logarithmic dependence of ToT on the number of p.e.\ is modelled with the fit function shown by the red line, which is used to estimate p.e.\ counts from raw experimental ToT values and correct for nonlinearity. %
The region between the dashed light blue lines represents the 1~$\upsigma$ uncertainty in the p.e.-to-ToT conversion, arising from uncertainties in the simulation parameters. The orange regions highlight the range of experimental ToT values obtained for muon events, showing that they predominantly fall within the region where ToT behaviour is closer to linearity. Specifically, the darker (lighter) orange region represents the range of ToT values that contribute 68\% (90\%) of the total p.e.\ in a muon event on average. The inset shows the 1 p.e.\ pulse measured from the TIA of PETsys ASIC and the fit used to characterise the average rise ($\uptau_r$) and decay ($\uptau_d$) times used in the simulation. The pulse shape is dominated by the response of the readout electronics---primarily the shaping circuitry of the TIA---while the intrinsic SiPM pulse shape has negligible impact on the overall signal. %
\label{fig:map_ToT_PEs}}
\end{figure}

In this work, we derive an analytical function describing the relationship between the ToT value and the number of p.e.\ that produced a signal. This correction function is developed by modelling the response of the SiPM and the electronic readout of the experimental setup, and by simulating SiPM pulses produced by a known number of p.e. %
To simulate SiPM pulses, we use SimSiPM, a dedicated library for SiPM simulations~\cite{simsipm}. %
The pulses generated by one p.e.\ and shaped by the PETsys Trans-Impedance Amplifier (TIA) are characterised by acquiring dark noise signals with an oscilloscope after the amplification stage of the debug channel of the PETsys readout. The rise and decay times of pulses generated by one p.e.\ dark noise are, on average, 6~ns and 48~ns, respectively. %
The experimental pulse and the fitted function used for its characterisation are included in the inset of Figure~\ref{fig:map_ToT_PEs}. SimSiPM is then used to simulate pulses generated by a known number of p.e. The arrival times of photons at the SiPM for each simulated pulse are modelled according to the scintillation decay times for LAB + 1~wt.\% PPO (5.5~ns (72\%), 13.7~ns (25\%), 84~ns (3\%)~\cite{Onken_2020}) and the Saint Gobain BCF-91A fibre decay time (8~ns, measured in a separate dedicated experiment). The additional POPOP in our scintillator is expected to have only a small effect on the decay times~\cite{Buck_2016}, and it has been shown that adding wax to make the scintillator opaque does not affect its scintillation kinetics~\cite{minipaper}. %
A dark count rate of 250~kHz and a crosstalk probability of 11\% are also set. %
The simulated pulses are analysed to determine how long they remain above a threshold. %
Several effects are considered in the simulation to mimic the electronics read out and account for systematic uncertainties. Channel-to-channel variations in the TIA output are included by randomly selecting the rise and decay times for each simulated signal within $\pm$15\% of the measured values. Additionally, the trigger threshold used to compute the ToT for each signal is randomly selected between 1.3 and 1.7 p.e.\ to account for small channel-to-channel gain variations. Nominally, the trigger threshold is set to 1.5 p.e.\ in the Cube detector, corresponding to the plateau observed in the acquisition rate versus trigger threshold curve between 1 and 2 p.e. %

Figure~\ref{fig:map_ToT_PEs} displays the relationship between calculated ToT and the number of p.e.\ generating the simulated pulse. %
Given the known logarithmic dependency of ToT as a function of the number of photoelectrons~\cite{sharma_2020}, a logarithmic function is used to fit the points in the graph, %
$\text{ToT}=\text{a}+\text{b}*\ln(\text{p.e.}+\text{c})$. %
The best parameters fitting the points are reported in the legend of Figure~\ref{fig:map_ToT_PEs}. The inverse function,
\begin{equation}
    \text{p.e.} = \text{e}^{\textstyle \frac{\text{ToT}+4.6}{52.4}} + 0.4,
    \label{eq:pe_tot}
\end{equation}
enables the estimation of the number of photoelectrons from the raw experimental ToT value and is applied throughout the remainder of this work to correct for nonlinearity. As highlighted by the orange regions in Fig.~\ref{fig:map_ToT_PEs}, the range of ToT values that contribute 68\% (darker) and 90\% (lighter) of the total photoelectrons in a typical muon event predominantly falls within the region where the ToT response is approximately linear. %
The conversion to p.e.\ corrects for the non-linearity in the raw ToT measurement and results are improved, as demonstrated in Section~\ref{sec:resolution}. %
The region between the dashed light blue lines in Figure~\ref{fig:map_ToT_PEs} represents the uncertainty associated with the p.e.-to-ToT conversion. It corresponds to the 1~$\upsigma$ range, arising from variations in the simulation parameters, with the shape of 1 p.e.\ pulses being the primary source of uncertainty. The conversion functions at the $\pm$1~$\upsigma$ extremes are further utilised in Section~\ref{sec:displays} in the discussion of the detector light level. %

\section{Muon light cylinders}
\label{sec:displays}
\begin{figure}[htbp]
\centering
\includegraphics[width=\textwidth]{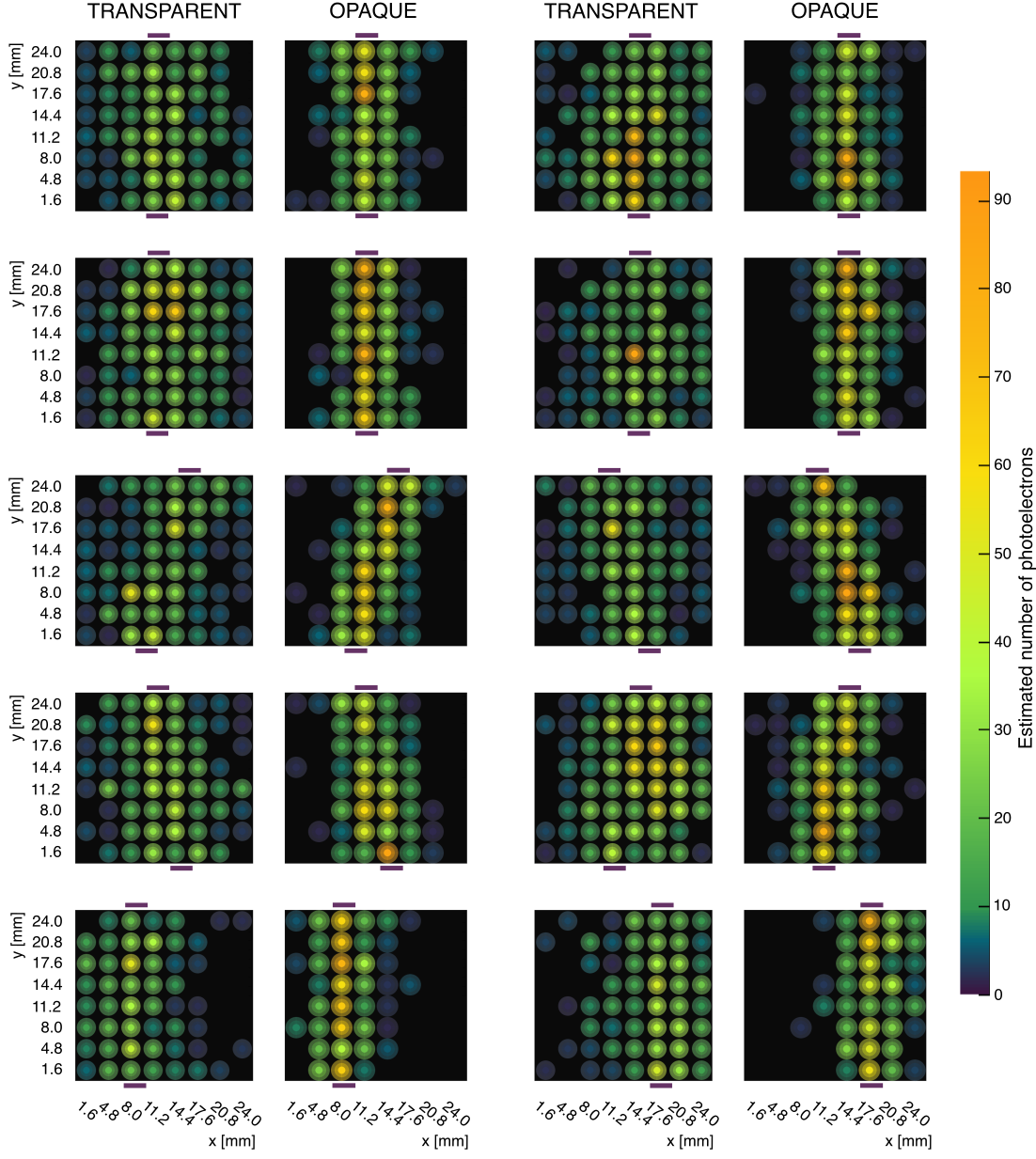}
\caption[Event displays]{Event displays of muons crossing the Cube. Each image represents the 8$\times$8 grid of WLS fibres and the colour scale shows the number of p.e.\ detected by each fibre. Events are grouped in pairs, selecting muons going through the same PixTags pixels in the transparent (left of pair) and opaque (right of pair) datasets. The size of the circles is exaggerated compared to the real diameter of the fibres for display purposes. The small purple rectangles above and below each image represent the Cube entry (top) and exit (bottom) location of the muon as estimated from the maximum signal in the top and bottom PixTags. %
\label{fig:evt_displays}}
\end{figure}

The stochastic confinement of light arising from the opacity of the scintillation medium can be seen in the maps of the light collected by each fibre in muon events. %
Figure~\ref{fig:evt_displays} presents example maps of the 8$\times$8 WLS fibre grid during a muon event from transparent and opaque scintillator datasets. The number of scintillation photons detected by each fibre is determined by summing the p.e.\ values (extracted from ToT using Eq.~\eqref{eq:pe_tot}) at both ends of the fibre. %

The small purple rectangles above and below each map indicate the entry and exit locations of the muon as estimated by the PixTags. These locations are derived from the projection along the line connecting the centres of the pixels with the highest signals in the top and bottom PixTags. The first event in time from both the opaque and transparent scintillator datasets involving the same pixels in the PixTags is selected and paired together in Figure~\ref{fig:evt_displays}. This allows for a side-by-side comparison of similar muon events when using the wax-based opaque scintillator (right display of each pair) and when the detector is filled with transparent scintillator (left display of each pair).

A clear distinction emerges between the two media. In the transparent case, scintillation light spreads broadly, leading to a more diffuse pattern of fibre signals. By contrast, the opaque NoWaSH scintillator confines light near its point of origin, resulting in more localised and trackable signal patterns. Since muons deposit energy continuously along their path, this confinement yields a series of overlapping, localised ``light balls'' that collectively form a distinctive ``light cylinder'' topology. This track-like structure is evident in the event displays of Figure~\ref{fig:evt_displays}, particularly in the right-hand panels of each pair, where the enhanced spatial confinement in the opaque medium leads to sharper, more defined tracks. %
\begin{figure}[htbp]
    \centering
    \begin{subfigure}[b]{0.45\textwidth}
        \centering
        \includegraphics[width=\textwidth]{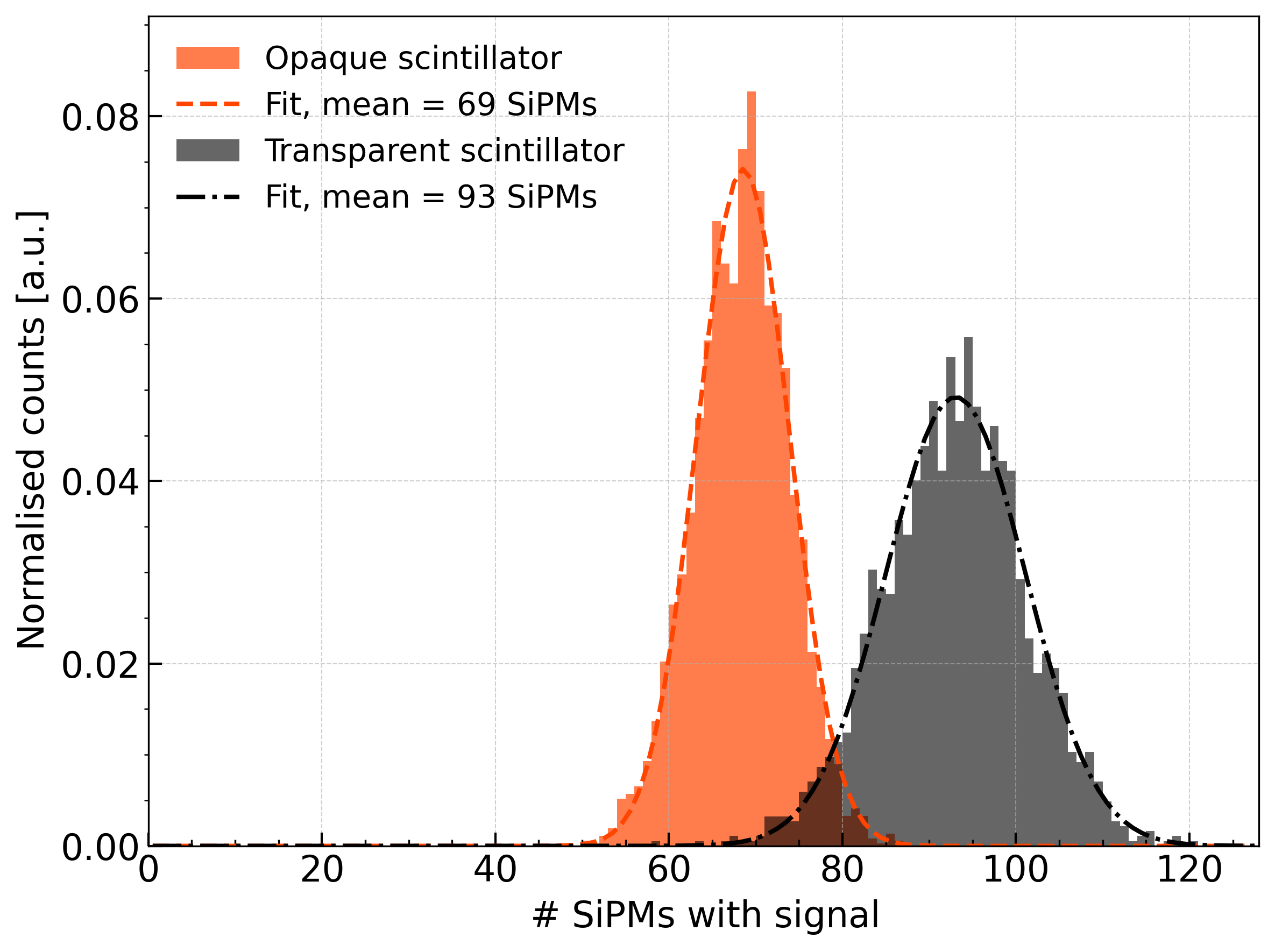}
        \caption{}
        \label{fig:distr_SiPMs_PEs/sipmsN}
    \end{subfigure}
    \hfill
    \begin{subfigure}[b]{0.45\textwidth}
        \centering
        \includegraphics[width=\textwidth]{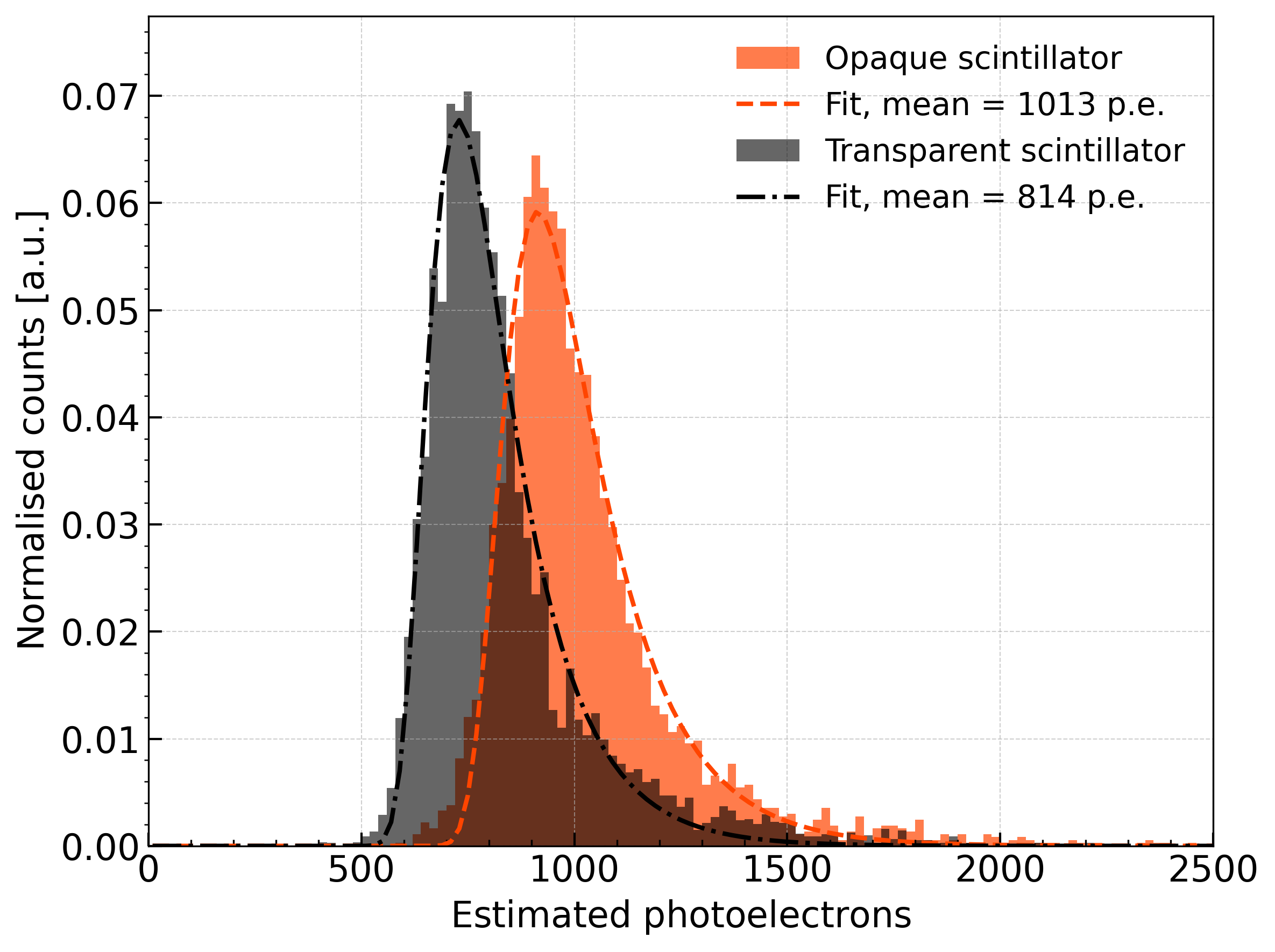}
        \caption{}
        \label{fig:distr_SiPMs_PEs/totEn}
    \end{subfigure}
    \caption[Distributions of the number of SiPMs with signal and of the total number of photoelectrons.]{%
    (a) Distribution of the number of SiPMs with signal and (b) the total number of photoelectrons in muon events. %
    The histograms correspond to data taken with the detector filled with either opaque (orange) or transparent (grey) scintillators. %
    A Gaussian function is used to fit the distribution of SiPMs with signal, while a Landau function is applied to the total number of p.e., as it is expected to describe the energy loss of cosmic ray muons. %
    When using the opaque scintillator, more light is detected and, furthermore, these photons are distributed across fewer SiPMs, demonstrating light confinement. %
\label{fig:distr_SiPMs_PEs}}
\end{figure}

This confinement effect is further supported by the comparison of the number of SiPMs with a signal in a muon event, as shown in Figure~\ref{fig:distr_SiPMs_PEs/sipmsN}. %
The number of SiPMs with signal per event is lower in the opaque case (average of 69) compared to the transparent case (average of 93), due to the confinement of the light from the muon track. %
The number of SiPMs hit was observed to be stable to 3\% over multiple weeks of data taking. %

Figure~\ref{fig:distr_SiPMs_PEs/totEn} shows the distribution of the total number of p.e.\ collected in each muon event for the two scintillators. About 25\% more light was collected when using the opaque scintillator, and furthermore, these photons were distributed across fewer SiPMs. The light level per row was reasonably uniform, with variations of approximately $\pm$10\%. The detected light can be estimated by considering that muons deposit approximately 2~MeV/cm and the height of the scintillator volume in the Cube is 3~cm, giving a total energy deposition of 6 MeV\@. In the opaque case, the average number of p.e.\ detected is about 170 p.e./MeV, while in the transparent is about 140 p.e./MeV\@. The uncertainty associated with these values of p.e./MeV is $\pm$40\% and is dominated by the conversion from ToT to p.e.\ described in Section~\ref{sec:tot}. %
Several factors limit the quantitative interpretation of the light levels in the two cases. For example, in the transparent setup wall reflectivity plays a more significant role in light loss than in the opaque case. Conversely, in the opaque case, the 15\% of the volume that is non-scintillating wax reduces the light yield compared to transparent. %

Overall, these results allow us to draw qualitative conclusions that provide insight for the optimisation of future LiquidO detector designs. Light confinement in LiquidO arises from two contributions: direct collection, in which the fibres themselves directly confine the light by stopping it from streaming away from its point of production, and stochastic confinement, given by the highly-scattering opaque medium. In the Cube detector the pitch of the fibres is 3.2~mm and their diameter is 1~mm, resulting in a relatively dense fibre lattice where the fibres occupy 7.7\% of the detector volume. This configuration positions the detector within a regime where light confinement results from both direct and stochastic contributions at comparable levels. As a result, muon tracks are observed with the transparent scintillator due to the direct confinement, albeit with significantly reduced sharpness compared to the opaque case. The relative contribution of stochastic confinement can be increased by several means: using thinner fibres, increasing the fibre pitch (i.e. reducing fibre density), or employing a medium with a shorter scattering length. However, these optimisations can come with trade offs in total light collection, spatial uniformity of light collection and timing resolution of the detector response, as well as practical considerations related to detector assembly and mechanical robustness. These design choices must therefore be carefully balanced in accordance with the specific performance goals of a given LiquidO detector application. %

\section{Tracking and position resolution}
\label{sec:resolution}
\begin{figure}[htbp]
\centering
\includegraphics[width=\textwidth]{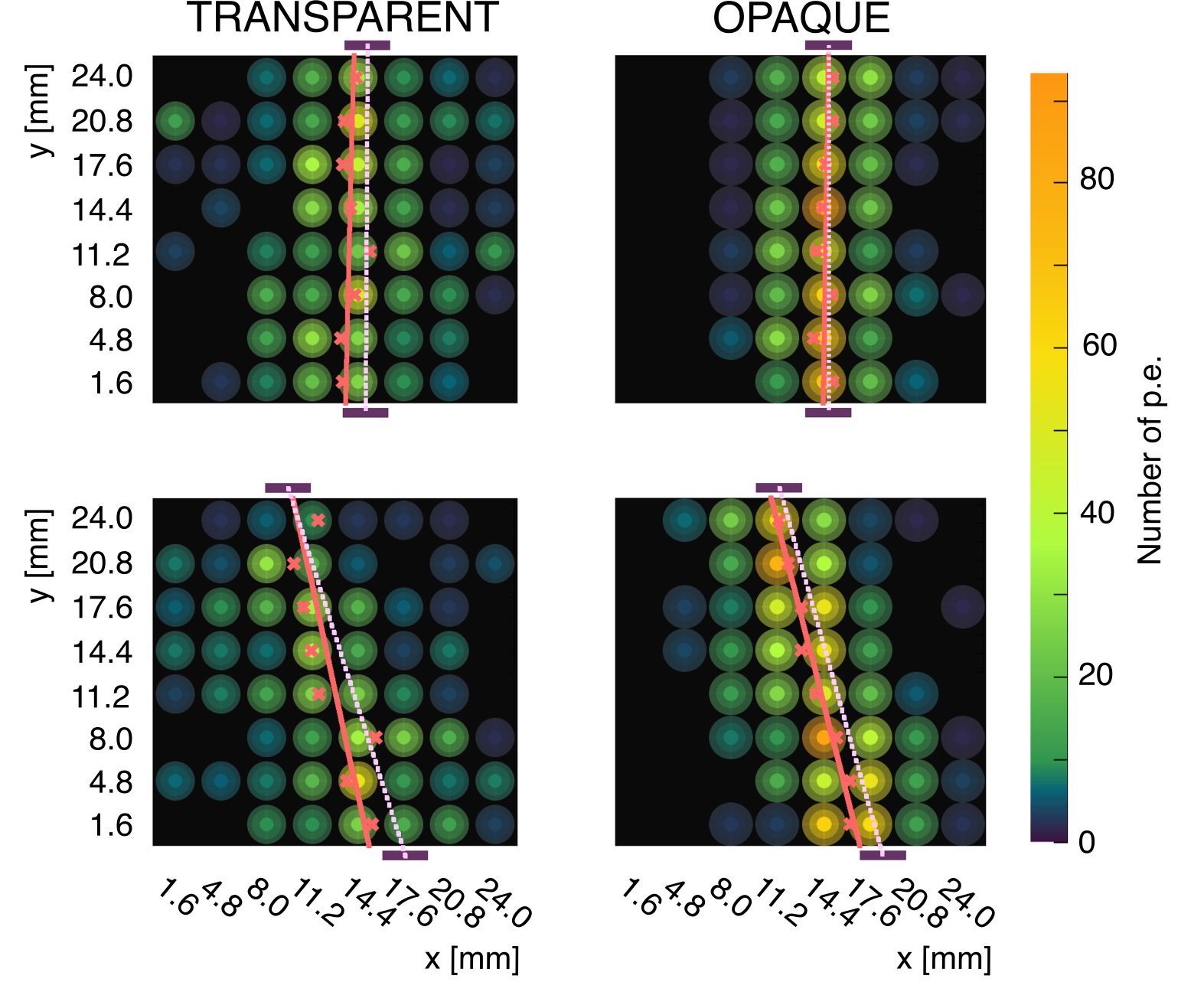}
\caption[Reconstruction examples.]{Examples of reconstructed muon positions (orange crosses) and tracks (orange line) in the Cube. The position is reconstructed for each of the eight rows of fibres. The tracks are produced by performing a straight-line fit to these positions. The tracks reconstructed using the information of the PixTags are also included (pink dashed line). Events are grouped in pairs, selecting muons going through the same PixTags pixels in the transparent (left of pair) and opaque (right of pair) datasets. %
\label{fig:reco_examples}}
\end{figure}

The accuracy of reconstructing muon tracks is crucial for evaluating the benefits of using an opaque scintillator. %
The position of a muon as it traverses the Cube is reconstructed for each of the eight fibre rows, using the hits recorded by the eight fibres in each row. The overall muon track is then determined by combining the spatial information from all rows. Specifically, for each row of fibres j, the reconstructed coordinate of the travelling muon $\text{X}^{\text{reco}}_{\text{j}}$ is calculated by a quantity analogous to the centre of mass with
\begin{equation}
    \text{X}^{\text{reco}}_{\text{j}} = \frac{\sum_\text{k} \text{w}_\text{k} \cdot \text{x}_\text{k}}{\sum_\text{k} \text{w}_\text{k}},
\label{eq:com}
\end{equation}
where $\text{x}_\text{k}$ is the x coordinate of the centre of fibre k and $\text{w}_\text{k}$ is a weight associated with that fibre.
The weights $\text{w}_\text{k}$ are a measure of the number of p.e.\ detected at the two ends of each fibre. The raw experimental ToT and the number of p.e.\ extracted using Eq.~\eqref{eq:pe_tot} are used as weights and the results are compared.
Reconstructed positions of muon tracks are shown in Figure~\ref{fig:reco_examples}, in the case of transparent and opaque scintillators. The tracks are produced by performing a chi-squared straight-line fit to the $\text{X}^{\text{reco}}_{\text{j}}$ points. %
This fit assumes muon trajectories are effectively straight within the detector, neglecting multiple scattering. Given the low atomic number and density of the scintillator, the short muon path length within the detector, and the high average energy of cosmic muons, this assumption is justified and any scattering-induced deviations are expected to be negligible. %
For comparison, tracks as would be inferred from only the PixTag signals are also shown in Figure~\ref{fig:reco_examples}. Note that information from the PixTags is used solely for muon selection and as a consistency check, and not for estimating the position resolution of the LiquidO detector. This is because the position information provided by the PixTags is significantly less precise than that of the Cube itself. %
From Figure~\ref{fig:reco_examples}, it can be observed that the tracks reconstructed by the Cube and the PixTags are broadly similar.

We investigate the position resolution, defined as the ability of each row of fibres to measure the position of a passing muon in the plane defined by fibres. The resolution of this reconstruction is measured by comparing the reconstructed position to a precise reference position of the muon, which is determined using the detector's inherent capability to track muon trajectories. Given that this approach relies on assessing the position resolution of a row of the Cube by considering a precise reference position estimated by the same detector, a method that corrects the final resolution for this effect is used~\cite{carnegie2005, arogancia2009}.

Considering the row j of the detector, the reconstructed position $\text{X}^{\text{reco}}_{\text{j}}$ is given by Eq.~\eqref{eq:com}. Two precise positions are then estimated. The first, $\text{X}^{\text{in}}_{\text{j}}$, is extracted using a linear fit to the $\text{X}^{\text{reco}}_{\text{i}}$ of every row of the Cube, including the row under test. The other, $\text{X}^{\text{ex}}_{\text{j}}$, is given by the linear fit through all the $\text{X}^{\text{reco}}_{\text{i}}$ excluding the row under assessment.
The distributions of the residual $\updelta^{\text{in}}_{\text{j}}=\text{X}^{\text{reco}}_{\text{j}}-\text{X}^{\text{in}}_{\text{j}}$ and $\updelta^{\text{ex}}_{\text{j}}=\text{X}^{\text{reco}}_{\text{j}}-\text{X}^{\text{ex}}_{\text{j}}$ are then generated and, finally, the position resolution is computed as
\begin{equation}
    \text{R}_\text{j} = \sqrt{\upsigma_{\updelta^{\text{in}}_{\text{j}}} \cdot \upsigma_{\updelta^{\text{ex}}_{\text{j}}}},
\label{eq:res}
\end{equation}
where $\upsigma_{\updelta^{\text{in}}_{\text{j}}}$ and $\upsigma_{\updelta^{\text{ex}}_{\text{j}}}$ are the standard deviations of the two residual distributions for the row j.
This approach has been shown to provide an unbiased estimate of the spatial resolution of a test detector when using other detectors with similar characteristics to reconstruct a precise reference position~\cite{alexopoulos2014}. That previous work is analogous to our study, where the ``detector'' under assessment is one of the rows of the Cube, while the ``reference detectors'' are the other seven rows.

\begin{figure}[htbp]
\centering
\includegraphics[width=\textwidth]{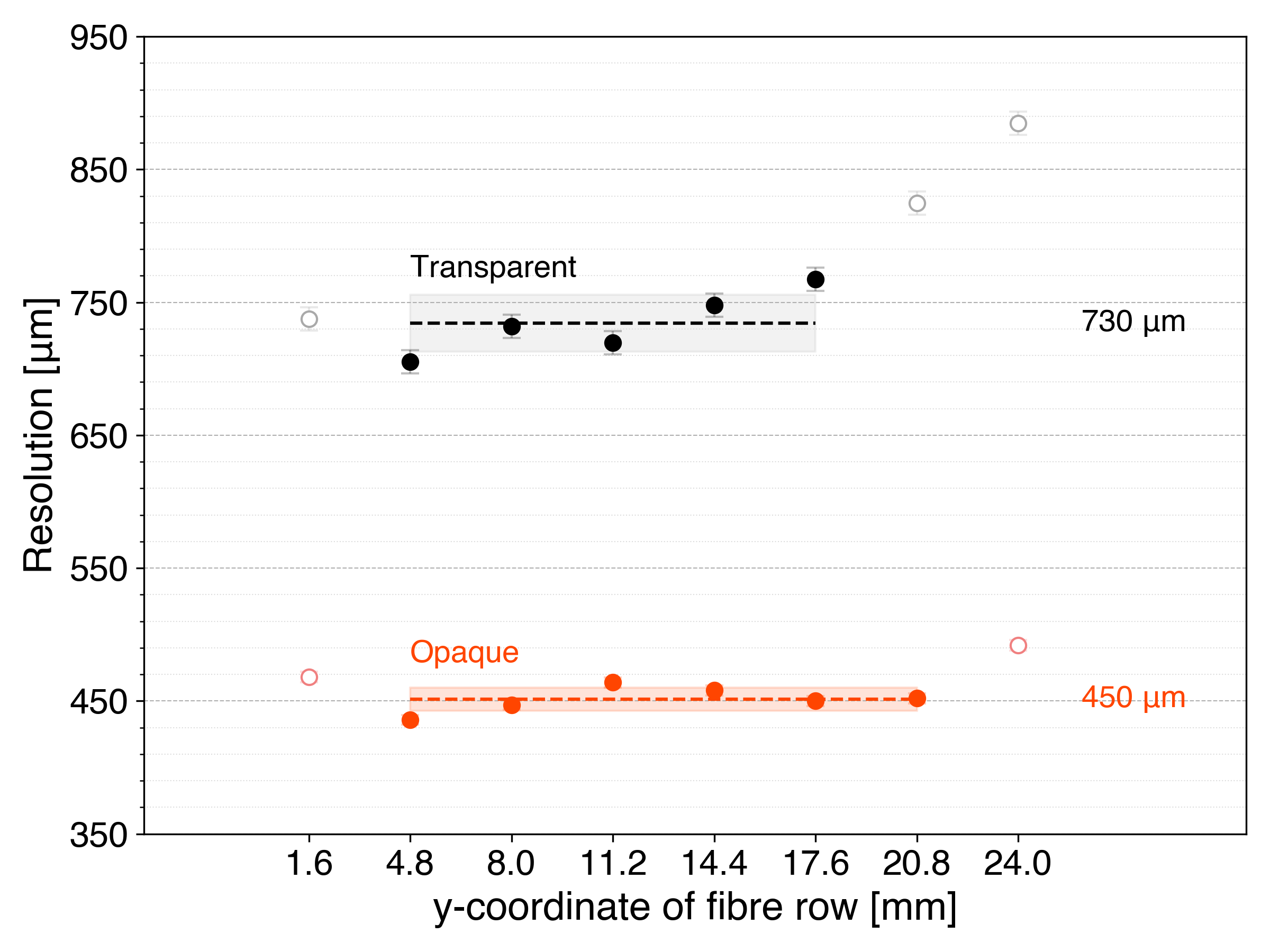}
\caption[Resolution results.]{%
Muon position resolution obtained for each of the eight rows of fibres in the Cube, filled with either transparent (black circles) or opaque (orange circles) scintillator. %
The reconstruction of the muon position uses the number of p.e.\ extracted from the raw ToT values. %
Filled circles represent the resolution for central rows, not impacted by edge effects, while open circles correspond to rows near the top (y $\geq$ 20.8~mm or y = 24~mm) and bottom (y = 1.6~mm) edges of the scintillator volume. The average resolution across the central rows is shown by the dashed lines. With the opaque scintillator, the average resolution per fibre row is 450~$\upmu$m, compared to 730~$\upmu$m in the transparent case---an improvement factor of 1.6. %
\label{fig:res_results}}
\end{figure}

The position resolutions for each of the eight rows estimated using this method are shown in Figure~\ref{fig:res_results}. Results for both the opaque and transparent scintillator are included. For each row, only the resolution calculated using the number of p.e., extracted from raw ToT experimental values, is shown. The row with y-coordinate 1.6~mm is the bottom one, as with the event displays shown in Figure~\ref{fig:evt_displays}.
The mean resolution is calculated using results from central rows that are not affected by border effects. These effects include reflections from the aluminium covering the top and bottom parts of the detector and possible air gaps between the reflective material and the scintillator at the top of the Cube. 
For the opaque case, to eliminate these factors, the top and bottom rows, closest to the reflective surfaces, are not included in calculating the average resolution. The position resolution of these rows is still reported as open circles in Figure~\ref{fig:res_results}, and it can be seen that it is slightly worse than that of the central rows.
In the transparent case, the impact of the proximity to reflective surfaces is greater, as light is not confined close to the emission point and is free to travel. %
The two top rows at y-coordinates 20.8 and 24.0~mm are less resolved because air gaps degraded the coupling between the scintillator and the aluminium reflective surface beyond the reflective border effect seen in the opaque scintillator. Therefore, these rows are not included in the average resolution calculation.

Each central row of the Cube detector filled with opaque scintillator resolved the passing muon's position to within 450~$\upmu$m on average, compared to 730~$\upmu$m when using the transparent scintillator---an improvement factor of 1.6. This enhancement is attributed to stochastic light confinement induced by the opacity of the material, which localises light near the point of energy deposition, as visually evident in the event displays of Figure~\ref{fig:evt_displays}. %
It should be noted that while the transparent setup's performance is considered final for the given fibre pitch, the opaque configuration has not yet been fully optimised. Further improvements in position resolution are expected, primarily through tuning the scintillator's opacity (i.e., scattering length), along with other design parameters. %

In both the transparent and opaque scintillator datasets, the position resolution is improved by 0.12~mm when the p.e.\ information is used, compared to raw ToT. %
This difference occurs because ToT does not scale linearly with the number of p.e., particularly within the core of the muon track where most light is collected. Converting ToT to p.e.\ mitigates this effect, enhancing the weights assigned to the core of the track and enabling more precise reconstruction. The uncertainty on the ToT-to-p.e.\ conversion discussed in Section~\ref{sec:tot} translates to minimal ($\pm$1\%) uncertainty on the resolution computed here. 

\section{Comparison with state-of-the-art muon-imaging scintillator-based detectors}
\label{sec:imaging}
A simple comparison of the position resolution performance can be made considering a hypothetical detector composed of transparent scintillator, which is optically and physically segmented with the same pitch as the Cube. When a muon passes through a row of segments in this hypothetical detector, a signal is generated only by the segment that the muon traverses. With a lateral size of the segment, or pitch, of L, this provides a resolution for the reconstruction of the muon position of L/$\sqrt{12}$, equivalent to the standard deviation of the uniform distribution. Therefore, for a 3.2~mm pitch, the resolution is 0.92~mm. %
The Cube detector with opaque scintillator achieves approximately two times better resolution than this segmented example, demonstrating that significant performance gains can be achieved with LiquidO, even prior to detector optimisation. %

A further comparison can be made considering state-of-the-art scintillator-based detectors in the field of Cosmic Ray Tomography (CRT). %
In CRT, spatial resolution is a key performance metric~\cite{PROCUREUR2018169}. For transmission-based imaging, accurate muon localisation enables the mapping of material density and is further improved by also reconstructing the muon direction. In scattering-based tomography, the deflection of muons by interactions with matter is measured, making angular resolution the critical factor~\cite{crt_ms}. To reconstruct the direction vector of the muon, several detectors with a planar geometry are typically stacked vertically; the achievable angular resolution depends on the position resolution of each layer and the spacing between them. %

Scintillators are widely used in CRT due to their robustness, ease of construction, cost-effectiveness, and high detection efficiency~\cite{crt_border_security, Bonomi_applications}. Modern muon imaging systems typically employ two or more double layers of orthogonal plastic bars to track particles in three dimensions. The scintillation light produced by the passing particles propagates directly to the bar end faces or is transported via secondary light emission in wavelength-shifting optical fibres. The light signal is read out by photodetectors optically coupled to the bar lateral edges. These bars can be easily shaped into various sizes, with the position resolution determined by their shape and lateral dimensions.

For rectangular bars, particles generally produce a single hit per plane, resulting in a spatial resolution of L/$\sqrt{12}$ as discussed above. This geometry is widely used because it is suitable for precise muon tracking and allows flexible, large-area detector configurations~\cite{RIGGI201816, borehole_1, Gluyas20180059, borehole_2}. A triangular shape can improve spatial resolution by measuring the signal fraction in adjacent bars, for example 2.5~mm resolution for $L = 15$~mm~\cite{luo2022lumis, anghel2015cript}. %

In addition to plastic bars, scintillating fibres are also used in muon tomography systems~\cite{gscan1, Mahon20180048, Chen_2023}. These fibres, when packed together, provide very fine segmentation equivalent to the fibre diameter, resulting in high spatial resolution. For instance, detectors using fibres with a diameter of 1~mm have achieved a resolution of 120~$\upmu$m by combining the information from multiple layers~\cite{gscan1}. %

LiquidO-based detectors offer clear advantages for muon imaging. The Cube prototype achieves a position resolution of 450~$\upmu$m at 3.2~mm pitch, 14\% of the pitch, suggesting that a 15~mm‑pitch LiquidO system could deliver around 2~mm resolution, compared to 2.5~mm in current triangular‑bar systems. This simple scaling indicates that LiquidO has the potential to match or outperform state‑of‑the‑art triangular-bar detectors and is expected to compete with, and potentially surpass, systems based on scintillating fibres. %

Further optimisation is needed to fully realise the capabilities of LiquidO detectors. Scaling to metre-scale devices introduces challenges such as light attenuation in longer fibres and maintaining scintillator uniformity across extended volumes. However, a design optimised for muon imaging could incorporate several enhancements: tuning key parameters---including scintillator scattering length, fibre diameter, and absorption length at emission wavelengths---and refining the detector geometry to maximise performance~\cite{josh_thesis}. For example, with the absorption lengths of our current scintillators, the best spatial resolution is expected to be achieved with a scattering length several hundred times shorter than the fibre pitch. Upgrading the readout electronics to improve photon-counting precision per SiPM would also be beneficial.
Crucially, unlike segmented systems where reconstruction is limited by sparse signal channels (typically one or two per detector layer), LiquidO provides richer optical information that enables more sophisticated reconstruction algorithms. In particular, machine-learning techniques can extract finer spatial features, correct systematic biases, and identify secondary structures such as $\updelta$-rays or overlapping tracks. Moreover, new formulations of opaque scintillator with higher light yield and wavelength-shifting fibres with higher trapping efficiency could enhance spatial resolution by increasing the amount of detected light, an improvement not equally accessible to segmented detectors. 
Collectively, these advances could improve position resolution by a factor of five to ten relative to a segmented detector of equivalent pitch. 

\section{Conclusion}
\label{sec:conclusion}
We have designed and characterised a 64-fibre LiquidO prototype, read out by 128 photosensors---the highest number employed in any LiquidO detector to date. Thousands of cosmic ray muons were externally tagged passing through the prototype, which was tested with both opaque and transparent scintillators. A comparison of images from the opaque and transparent datasets demonstrated, on an event-by-event basis, the significantly enhanced light confinement achieved using the wax-based, highly-scattering scintillating medium. This provides a clear demonstration of the LiquidO effect.

The tracking and position reconstruction capabilities were also evaluated. Specifically, individual rows of fibres with a 3.2~mm pitch gave a position resolution of 450~$\upmu$m when using the opaque scintillator. In comparison, the resolution achieved is twice as precise as that expected for a simple segmented detector with the same pitch, demonstrating the superior imaging performance of LiquidO technology in resolving track-like events.

These results highlight the potential of LiquidO-based detectors for applications requiring millimetre and even sub-mm position reconstruction. In the context of muon imaging, they demonstrate the ability to compete with and potentially surpass current state-of-the-art scintillator-based technologies, whether in terms of spatial resolution, cost-effectiveness, or design flexibility. %

\acknowledgments
We gratefully acknowledge the support in the UK by STFC (grants ST/S000798/1, ST/W005700/1, ST/V001361/1, ST/W000512/1, ST/X006026/1, ST/X004961/1) and the Royal Society (grants SRF/R1/231053, RSWVF/R3/223011), in Germany by the Cluster of Excellence ``Precision Physics, Fundamental Interactions, and Structure of Matter'' (PRISMA$^{+}$ EXC 2118/1) funded by the German Research Foundation (DFG) within the German Excellence Strategy (Project ID 390831469). %
We gratefully acknowledge the support of the CNPq/CAPES in Brazil, the McDonald Institute providing FVRF support in Canada, the Charles University in the Czech Republic, the CNRS/IN2P3 in France, the INFN in Italy, the Fundação para a Ciência e a Tecnologia (FCT) in Portugal, the CIEMAT in Spain, the University of California at Irvine, Department of Defense, USA, Defense Threat Reduction Agency, USA (HDTRA1-20-2-0002) and the Department of Energy, National Nuclear Security Administration, Consortium for Monitoring, Technology, and Verification (DE-NA0003920), Brookhaven National Laboratory supported by the U.S. Department of Energy under contract DE-AC02-98CH10886, U.S. National Science Foundation-Major Research Instrumentation Program under contract PHY-2018280, Deep Learning for Statistics, Astrophysics, Geoscience, Engineering, Meteorology and Atmospheric Science, Physical Sciences and Psychology (DL-SAGEMAPP) at the Institute for Computational and Data Sciences (ICDS) at the Pennsylvania State University in the USA for their provision of resources. %

\section*{Authors Contributions}
The Cube prototype detector project was led by the University of Sussex team, with significant hardware contributions from JGU Mainz / PRISMA$^+$. %
A.G-F. constructed the setup, with ideas and support from A.E\@. %
The commissioning of the setup involved collaborative efforts from A.E., E.F.B., J.A.L., J.C.C.P., T.J.C.B., and W.S\@. %
Early versions of the simulation were developed by G.W. and B.J.C., while W.S. and N.T. investigated the optimisation of LiquidO detectors. %
Operations and collection of the data shown here were managed by J.A.L. and N.T\@. %
The analysis methodology was developed by N.T.\ and cross-checked by M.G\@. %
Results were produced by N.T. and verified by J.A.L\@. %
The short-scattering length opaque scintillator was developed and supplied by S.S\@. %
The publication write-up was a joint effort by J.H., J.A.L., and N.T., with all authors contributing their expertise in LiquidO technology to the final manuscript. %





\bibliographystyle{JHEP}
\bibliography{biblio}

\end{document}